\documentclass[a4paper]{jpconf}
\usepackage{graphicx}
\usepackage{wrapfig}

\begin{document}

\title{Improving Calibration of the Large Low-Background Proportional Counter}

\author
{O.D.~Petrenko$^{2}$, A.M.~Gangapshev$^{1}$, Yu.M.~Gavrilyuk$^{1}$, V.V.~Kazalov$^{1}$,
V.V.~Kuzminov$^{1}$, S.I.~Panasenko$^{2}$,
S.S.~Ratkevich$^{2}$,
D.A.~Tekueva$^{1}$
}
\address{$^1$ Baksan Neutrino Observatory INR RAS, Russia}
\address{$^2$ V.N.Karazin Kharkiv National University, Ukraine}

\ead{petrenko1694@gmail.com }

\begin{abstract}
The study of low-yield effects requires not only good quality of the original data but
also puts high requirements for their processing procedures to increase the efficiency
of the selection of useful events. The exploiting of the large cylindrical proportional
counter's electrostatic topology allows improving the extrapolation of information
about the primary ionization of a multipoint event. Long-term calibration measurements
with an external $^{109}$Cd-source allowed the development of a new method for
analyzing the pulse shape from a sizeable proportional counter. Optimized analysis of
the current's pulse shape from the electron cloud of primary ionization in the counter
improved the resolution and energy calibration. As a result, the efficiency of
selecting useful events was increased by 25\%.
\end{abstract}

\section{Introduction}
The search for such a rare nuclear decay as the simultaneous capture of two orbital $2K$
electrons by the same nucleus using large proportional chambers, in addition to a low
intrinsic background, requires unique working properties of the detector, such as the
stability of spectrometric characteristics overtime during very long measurements.
Recording the response of the relaxation of atomic processes after the capture of orbital
electrons in a gaseous medium provides essential advantages over liquid and solid-state
detectors. The interaction of radiation with gas atoms makes it possible to reveal the
topological signature of a rare event, which was demonstrated in our previous works
\cite{rf1,rf2a}.
When the source is a part of the detection medium, it is possible to measure the total
energy release except for a neutrino emission.
In the case of using solid-state or liquid detectors, the atomic shell's relaxation
products have sufficiently low energy, and the capture event of $2K$ is practically
one-point.
It has no distinguishing features compared to background events of the same energy.
Therefore, the search for the effect is carried out by the excess of the peak over the
full background in a given energy range, as was demonstrated by the XENON collaboration
\cite{XENON1}.
In the case of studying $2K$ capture in isotopes of gaseous xenon, there are additional
advantages such as much smaller fluctuations in the formation of an ionization charge
compared to the liquid state. This leads to a better intrinsic energy resolution of the
ionization signal by more than an order of magnitude \cite{Boltnikov97}.

In a gaseous medium, a characteristic quantum can travel rather long distances from birth
to absorption. In an event with the emission of two characteristic quanta absorbed in the
working gas and one Auger electron, the energy will be distributed in three point-like
regions. It is these events with a unique set of features that are the subject of our
research.

Such a rare phenomenon as a double-$K$-shell photoionization of the
atom can create the ``hollow atom'' by absorbing a single photon and releasing both $K$
electrons. Detection of such a process is possible by observing double-$K$-satellite
fluorescence transitions during relaxation of these states.
This process can be a source of background in an experiment to search for $2K$ capture of
$^{124,126}$Xe and, at the same time, serve as a methodological test for analyzing the
accumulated data.

This report is devoted to both the analysis of changes in the spectrometric properties
overtime of a xenon detector during long-term measurements and attempts to increase the
efficiency of the selection of useful events in the detector.

\section{Calibration of the energy spectrum}

Experiments on studying the processes of double beta decay with the capture of electrons
of an atom's inner shell in various isotopes of nobel gases using several 10-liter copper
proportional counters (CPC) have been carried out for a long time at the Baksan Neutrino
Observatory of the INR RAS.
The experience of long-term measurements to detect the capture of $2\nu2K$ at samples of
different enrichment in $^{78}$Kr and $^{124,126}$Xe using large proportional counters
with a copper body showed that the spectrometric properties of the detector noticeably
deteriorate with time.
We assume that this is due to the design features of detectors made of electrolytic
copper, which is characterized by a reduced content of radioactive impurities, but still
has internal microscopic particles of residual oxygen.

\begin{wrapfigure}[28]{l}{0.5\textwidth}
\includegraphics[width=19.0pc,angle=0]{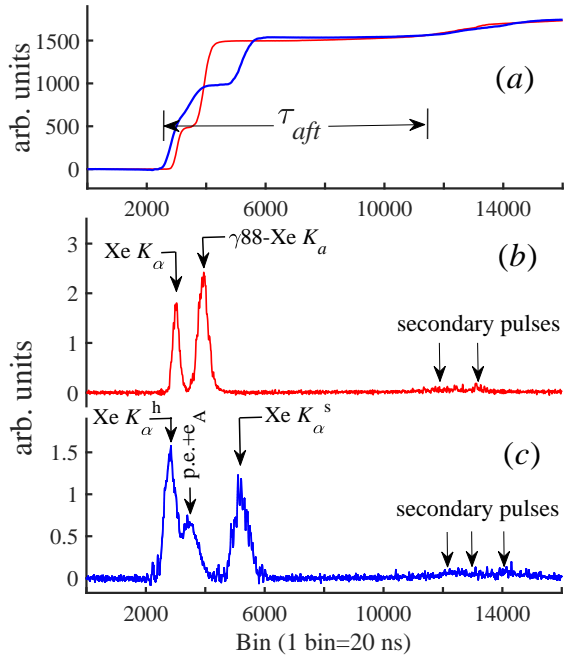} \vspace{-1pc}
\caption{\label{pic1} {\small ({\it a}) - examples of charge pulses from CSA corrected
for self-discharge from two- (red curve) and three-point events (blue curve).
Current pulses for two- and three-point events, $(b)$ and $(c)$ respectively.
$\tau_{aft}$ is the duration of the electron drift from the cathode to
the anode.
}}
\end{wrapfigure}

This study aims to determine the change in the detector's spectrometric properties under
prolonged exposure to gamma radiation from an external source of $^{109}$Cd when the CPC
is filled with xenon (or krypton) under a pressure from 4.2 to 5.0 bar.
A $^{109}$Cd isotope decays into $^{109}$Ag with a half-life of
461 days, emitting a number of ${\gamma}$-ray lines in the
energy range of 20-90 keV. The source was produced by the Cyclotron company and had an
activity of $0.238$ GBq in December 2015 \cite{Cd109}.
The 88-keV ${\gamma}$-rays with a relative yield of 0.036 photons per decay of $^{109}$Cd
passed through a collimating hole located in the middle of the detector's length and
irradiated the working medium of the counter through the wall of its casing.
The fundamental difference between low-background nuclear physics measurements and other
types of studies of ionizing radiation sources' characteristics lies in the shallow
overall event counting rate. Therefore, it is possible to register any single response of
the detector and, after digitizing with a high frequency of the charge-sensitive
amplifier's (CSA) output pulses in the offline mode, analyze waveform recorded for each
event. In our measuring the signals were digitized with 12 bits precision and 50 MHz
sampling frequency. 16K samples were recorded per pulse (Fig. \ref{pic1}{\it a}).
After a $\sim50$ $\mu$s long baseline the charge signal rises upwith from 10 to 40 $\mu$s
rise time followed by a $\sim200$ $\mu$s long exponential tail due to the discharge of
the feedback capacitor.
The CSA charge pulses for multipoint events look like a pile of separate signals.

Figure \ref{pic1}{\it a}
shows examples of charge pulses for the two-point event (photoabsorption - red curve) and
three-point cases due to double photoionization on the $K$ shell of the xenon atom (blue
curve). The pulses were obtained after correction on an exponential self-discharge CSA
and subtraction of the ionic component (see \cite{PTE}).
The current pulses corresponding to the electron component in pointwise charge clusters
of primary ionization in gas for two-and three-point events illustrated in
Fig.\ref{pic1}{\it b} and 2{\it c} respectively.
The obtained shape of the current pulse can be described by a set of Gaussian curves that
enable us to determine the charge brought in a certain component of the multipoint event.
Calculated area of an individual Gaussian should correspond to the charge (energy) of the
corresponding pointlike ionization.
As seen from the figure after a while the main pulse the secondary afterpulses follow.
They are because noble gases present high electroluminescence yields, emitting mainly in
the VUV region.
During a Townsend avalanche, the electroluminescence gives useful information for the event's tracking along the length's anode wire.
The photons of electroluminescence knock out low-energy electrons from the copper cathode.
Their amplitude is several times less than the amplitude of the main pulse, and the shape
duplicates the shape of the primary pulse. The photoelectric effect at the cathode is
quite probable since there are no quenching additives in the working gas. The delay
between pulse and afterpulse ($\tau_{aft}$) is determined by the total drift time of
electrons from the cathode to the anode. It sets the duration of the time interval in
which any single event is fully contained, regardless of the distribution of primary
ionization over the CPC volume.
The ratio of the amplitude of the secondary signal to the amplitude of the primary signal allows one to select events with a required coordinate, thus excluding the edge responses of CPC \cite{Nucl2010}.

\begin{figure}[h]
\begin{center}
\includegraphics[width=30.0pc,angle=0]{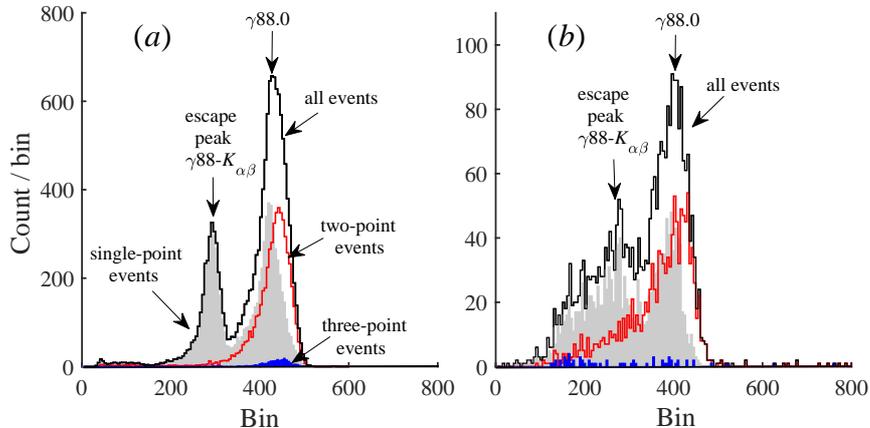}
\hspace{-4pc}
\caption{\label{pic2} {\small Pulse height spectrum from the external $^{109}$Cd source
located in the middle of the CPC length. {\it a} - the CPC is filled pure Xe with total
pressure of 5.0 bar;  no quenching or accelerating gases are added.   {\it b} - the same
sample, but containing micro additions of oxygen and hydrogen sublimated from the inner
surfaces of the detector walls after 4000 hours of counter operation.
}}
\end{center}
\end{figure}

As see, the bell-shaped distribution of the current signal allows for more confident
detection of multipoint events.
The area of the peak in the current pulse is proportional to the number of electrons of
primary point ionization.
The spectra of one-, two-, and three-point events in Fig.\ref{pic2} were constructed from
the total peak areas in the primary current signal taking into account the pulse shape
discrimination.

For pure xenon, the drift time of ionization electrons from the cathode to the anode is
$\sim160$ $\mu s$ at a pressure of 5 bar.
The number of electrons of primary ionization can be estimated on the area under
bell-shaped distribution (or the sum of the areas in the case of a multipoint event) in a
current pulse in the time interval
from the beginning the primary pulse up to the afterpulse.

Figure \ref{pic2}
shows the amplitude spectra of the total energy releases in the counter (``all events''),
obtained, respectively, immediately after gas purification and at the end of the run
after 4000 hours.
Before filling of the counter, working gas was purified from electronegative impurity
(O$_2$ and N$_2$) by flushing the gas through a Ti reactor at 800$^\circ$.
Gas contaminants with electronegative molecules lead in signal reduction, degradation of
the energy resolution and background discrimination capabilities.
Their effect particularly pronounced in regions of the low electric field.

The energy resolution $R$ of the peak at 58.2 keV when the characteristic radiation
escape from the gas without further interaction and the 88 keV line determined from the
peaks' right-hand halves turned out to be 10\%
and 12.5\%,
respectively.
The relatively low resolution and deviation from the dependence $R(E) = const / \sqrt{E}$
is explained by the fact that the counter operation mode corresponds to the beginning of
the limited proportionality region for pressure 5 bar due to particular needs.

The limited proportionality leads to weak nonlinearity of the detector response in
addition to the linearity jump of $177\pm9$ eV at the $K$-edge of absorption in xenon at
34.6 keV \cite{Tsuneini93}
Thereby, the centroids of the 88-keV peaks for one-point and two-point events do not
coincide.

The low-energy part of the spectrum is created by characteristic radiation ($E \approx
22-25$ keV) associated with transitions in $^{109}$Cd$\rightarrow^{109}$Ag, which
``survived'' after passing through a 5-mm-thick copper wall; scattered radiation from the wall, which is in equilibrium with the characteristic radiation, and Compton electrons
from a scattering of 88-keV photons in the gas with the escape of a Compton photon from
the active volume of the detector.

\section{
Spectrometric performance of the CPC at long-term measurements}

Comparing the spectra of total energy releases at the beginning and the end of the run,
one can see the degradation of the spectrum over time.
At the same time, it was noticed that the parameters of signals from the CPC for xenon
and krypton fillings behave differently in time.
\begin{figure}[h]
\begin{minipage}{17.5pc} \hspace{-1pc}
\includegraphics[width=18pc]{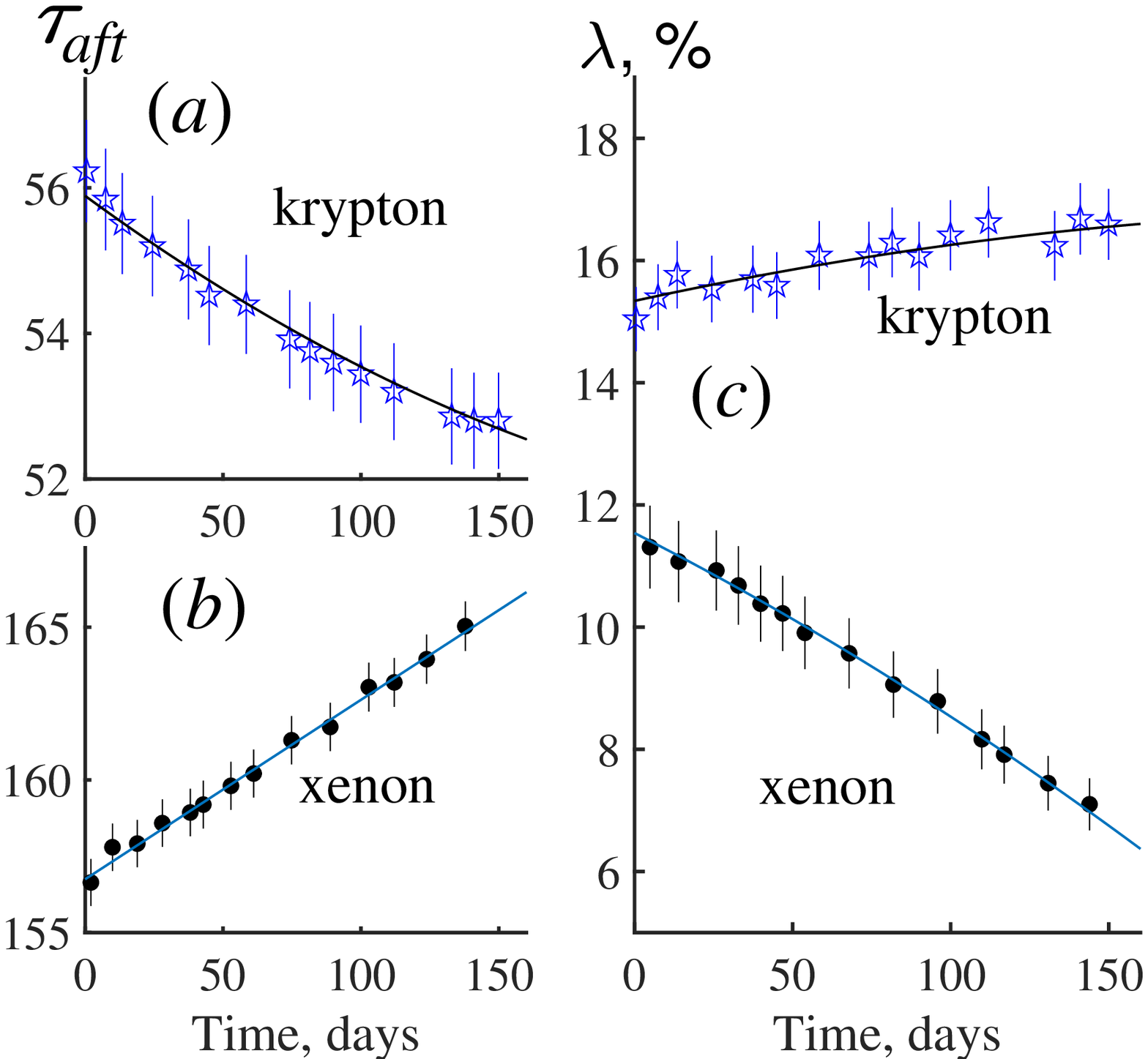}
\caption{\label{pic3} The variation of electrons drift time
from the cathode to the anode $(\tau_{aft})$ and the ratio of the amplitudes of the
afterpulse to the primary pulse $(\lambda)$ for long-term measurements at filling of the
CPC with krypton and xenon performed in the same conditions.}
\end{minipage} \hspace{1pc}%
\begin{minipage}{18.5pc} \vspace{-1.5pc}%
\hspace{-0.5pc}
\includegraphics[width=19pc]{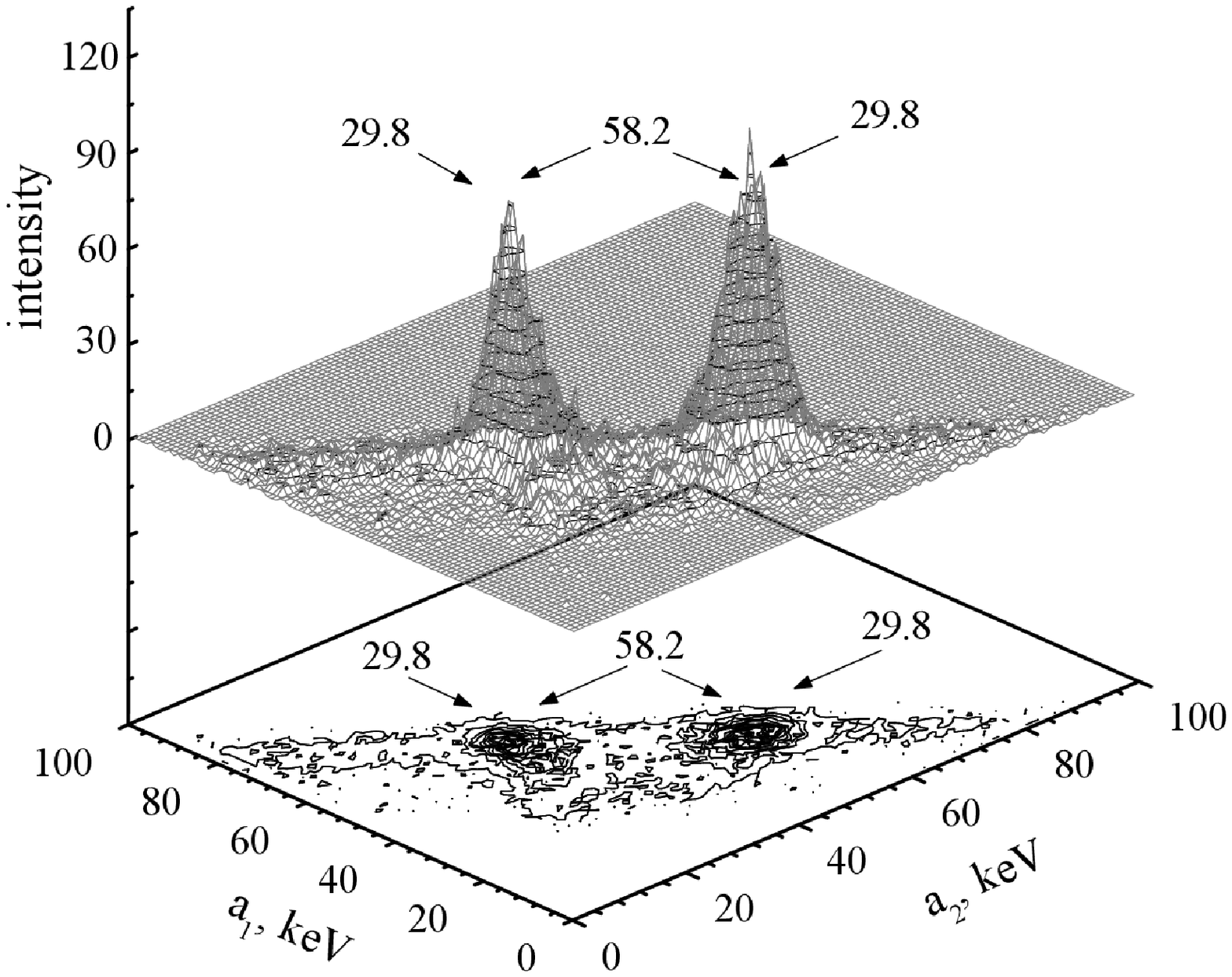}
\caption{\label{pic4} The distributions of amplitudes of energy depositions of individual
components for two-point events obtained by irradiation of a xenon-filled CPC with 88-keV
photons.}
\end{minipage}
\end{figure}
Figure \ref{pic3} demonstrates the change in the drift time of the knocked out electrons
from the cathode to the anode and the relative degradation of the afterpulse amplitude.
The temporal parameters of an individual pulse changed slowly. So the delay between the
pulse and afterpulse, which depends on the total drift time of electrons from the cathode
to the anode, increased for xenon, while for krypton, it decreased.
As can see, the electron drift velocity decreases over time for xenon, compared to the
krypton filling.
The behavior of the ratio of the amplitudes of afterpulses to the main pulse is also
different.
Two factors can explain such behavior of these parameters over time for long-term
measurements with the help of the large cylindrical counter: the slow sublimation of
polyatomic gas molecules originating from the surfaces of the case and insulators;
besides, oxygen diffuses gradually from the flange sealing rings of the CPC.
\begin{figure}[h]
\begin{minipage}{18pc} \hspace{0pc}
\includegraphics[width=18pc]{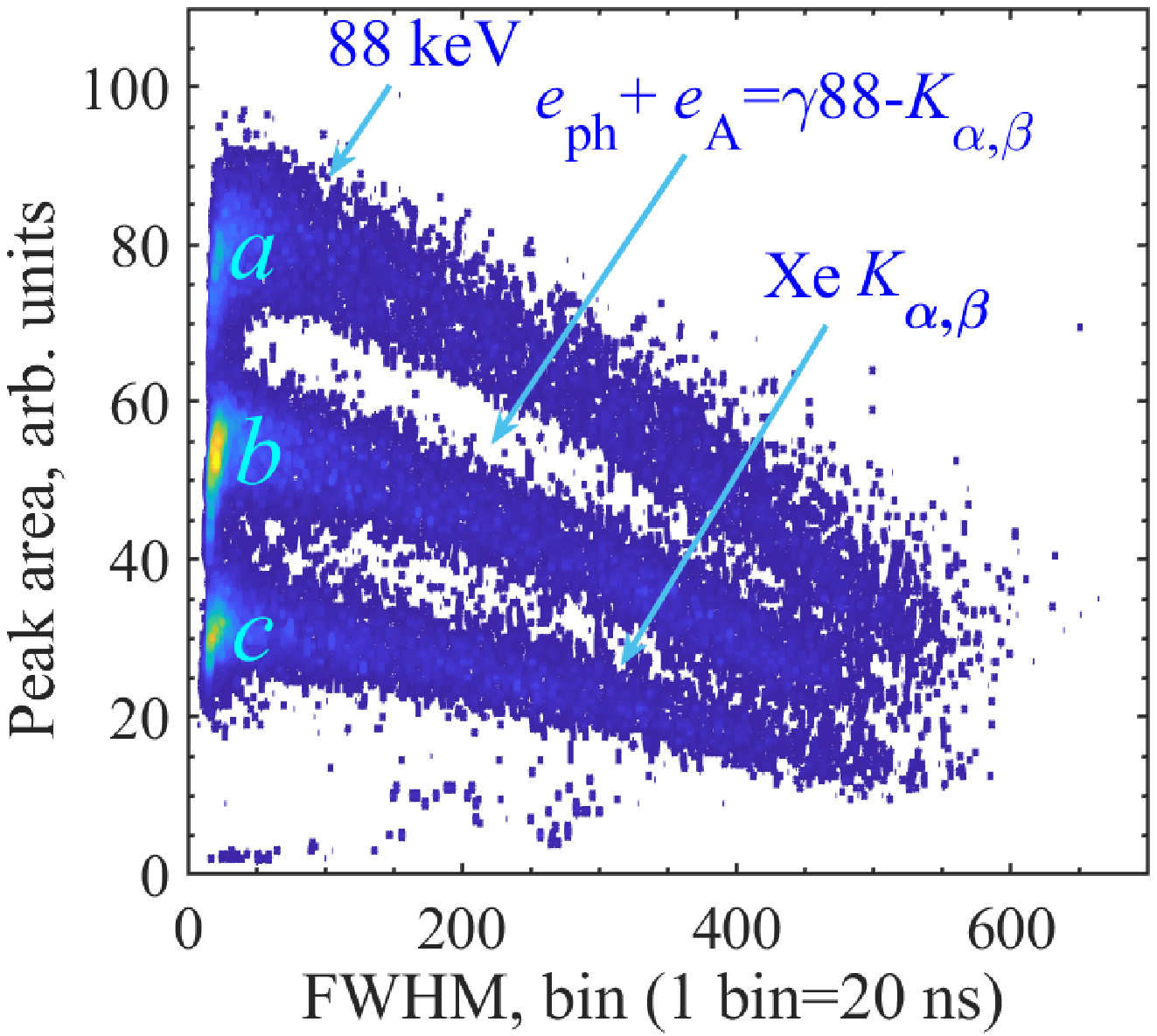}
\caption{\label{pic5} The two-dimensional distribution of points depending on the area
and width of the current signal peak.}
\end{minipage} \hspace{3pc}%
\begin{minipage}{17pc} 
\includegraphics[width=15pc]{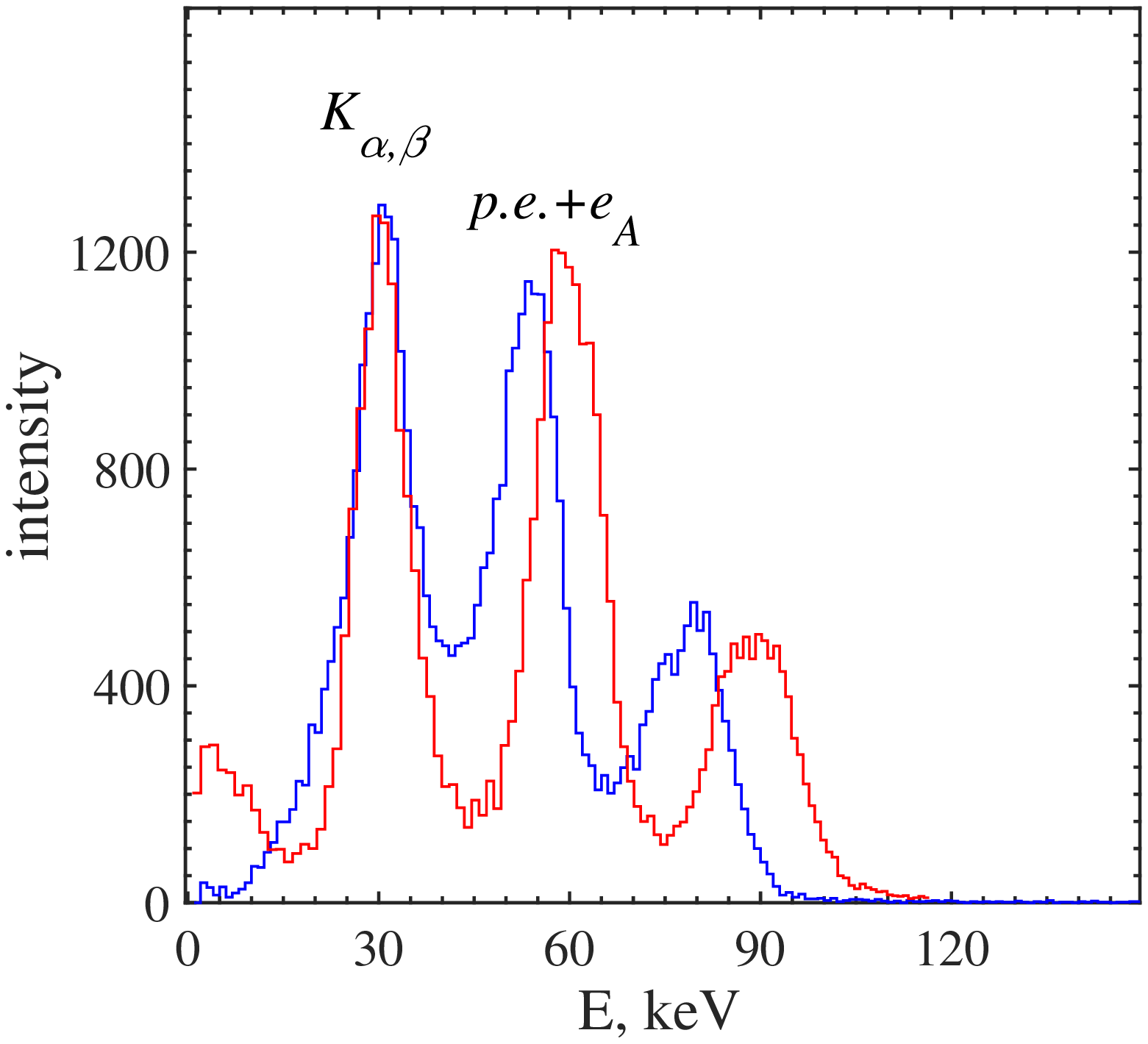}
\caption{\label{pic6} {\small
The initial (blue curve) and corrected spectra (red curve) of individual single-point charge clusters of primary
ionization.
}
}
\end{minipage}
\end{figure}

Fluorescence X-rays are a product of cascade decay of vacancies in a photoionized atom
and can represent a significant part of the characteristic response of a gas detector.
The likelihood of fluorescent X-ray emission will depend on the volume of the detector,
the filling gas and pressure, and the energy of the fluorescent radiation.

The photoeffect on the $K$ shell of xenon is 83.8\%
and the one on other shells is 16.2\%
\cite{xdb}.
Characteristic radiation is emitted in 89.4\%
when a vacancy in the $K$ shell of xenon is filled, and Auger electrons are in 10.6\%.
As a result, in 74.9\%
of photoabsorption events are formed with the emission of a $K$-photon
and the sum of electrons $(e_{ph}+e_A)$ with a high degree of probability creates two
pointwise charge clusters spaced at a sufficiently large distance, that generate two
bell-shaped current signals (see Fig. 1).
The calculated absorption efficiency of characteristic radiation in the working volume of the CPC is 0.735.
Consequently, 55.1\%
of two-point events are initially created by the photoelectric effect at the peak of
total absorption.
Using the pulse shape discrimination of a current signal allows us to select the
two-point events with high efficiency.

Figure \ref{pic4} shows the intensity distribution of two-pointwise charge clusters
depending on the amplitude of the first $(a1)$ and second $(a2)$ subpulses in the current signal. Intensity maxima are observed for a combination of subpulse amplitudes
($K_\alpha=29.8$ keV; $E_{ph.e.+e_{A}}=58.2$ keV).
A diagonal stripe is seen that includes events with a total absorption in the range from
60 to 90 keV created by the photoelectric effect with emission $e_A(K)$ and
bremsstrahlung photons, as well as by Compton scattering.
Reliable separation of responses from primary ionization in the detector in the case of
$K$ X-rays and the case of electrons makes it possible to indirectly investigate the
blurring of the charge spot as a function of the drift time from the place of track
formation to the anode.

So, from the detector's collection of signals, we can isolate point ionization from $K$
X-ray, photoelectrons, and Auger electrons.
Figure \ref{pic5} shows a two-dimensional distribution of points depending on the area
and width of the current signal peak.
The figure clearly shows three extensive groups of point loci.
The first group $(a)$ is formed from the responses of the CPC by 88 keV line mainly due
to the photoelectric effect on the outer shells of the xenon atom and the sum of $K$
X-rays and a photoelectron that is spatially not resolved in projection onto the anode.The second group events $(b)$ form during the registration of a photoelectron with $E_{ph.e.}=\gamma 88 - K_{ab}$ and an Auger electron in a single pointwise charge cluster of primary ionization when the $K$-fluorescence X-rays escape from the active volume of the detector.
The third group $(c)$ displays responses caused by
xenon characteristic $K$ lines with the energy $\sim 30$ keV.
It is clearly seen how the amount of registered charge corresponding to the number of
electrons that reached the anode from ionization by a soft photon or photoelectron of the working gas depends on the duration of the current burst.
The wide signals can be explained by events born near the cathode.
The primary ionization electrons for such events have the maximum drift time to the anode and have a more significant number of recombinations.

Based on these distributions, it is possible to determine the loss of primary electrons
due to recombination depending on the drift time of the initial charge swarm.
We describe the dependence of the area and width of the bell-shaped current signal by a
polynomial of the third degree.
This allows us to introduce a correction to the energy release spectrum in the detector.
The initial and corrected spectra of all events from the three groups listed above are
shown in Fig.~\ref{pic6}.

\section{Conclusions}
The proposed method of analysis of the data of calibration measurements with the $^{109}$Cd source allows one to increase the accuracy of the energy determination and to improve spectra resolution obtained with a large proportional counter.
The response correction improved the energy resolution $(\sigma / E)$,
from $27 \pm 0.1 $
to $16.0 \pm 0.3 $ percent
for 30 keV and increased the efficiency of useful event selection up to 25\%.

\end{document}